 \documentclass[final,5p,times,twocolumn]{elsarticle}

\usepackage{amssymb}
\usepackage{color}
\usepackage[normalem]{ulem}
\biboptions{sort&compress}

\journal{Fundamental Research}

\begin{document}

\begin{frontmatter}



\title{Parity-dependent unidirectional and chiral photon transfer in reversed-dissipation cavity optomechanics}


\author[inst1]{Zhen Chen}

\affiliation[inst1]{organization={Beijing Academy of Quantum Information Sciences}, city={Beijing}, postcode={100193}, country={China}}

\author[inst1]{Qichun Liu}
\author[inst2]{Jingwei Zhou}

\affiliation[inst2]{organization={CAS Key Laboratory of Microscale Magnetic Resonance, University of Science and Technology of China}, city={Hefei}, postcode={230026}, country={China}}

\author[inst1]{Peng Zhao}

\author[inst1]{Haifeng Yu\corref{cor1}}
\ead{hfyu@baqis.ac.cn}

\author[inst3,inst1]{Tiefu Li\corref{cor1}}
\ead{litf@tsinghua.edu.cn}
\affiliation[inst3]{organization={School of Integrated Circuits and Frontier Science Center for Quantum Information, Tsinghua University}, city={Beijing}, postcode={100084}, country={China}}

\author[inst1]{Yulong Liu\corref{cor1}}
\ead{liuyl@baqis.ac.cn}
\cortext[cor1]{Corresponding author}
\begin{abstract}
Nonreciprocal elements, such as isolators and circulators, play an important role in classical and quantum information processing. Recently, strong nonreciprocal effects have been experimentally demonstrated in cavity optomechanical systems. In these approaches, the bandwidth of the nonreciprocal photon transmission is limited by the mechanical resonator linewidth, which is arguably much smaller than the linewidths of the cavity modes in most electromechanical or optomechanical devices. In this work, we demonstrate broadband nonreciprocal photon transmission in the \emph{reversed-dissipation} regime, where the mechanical mode with a large decay rate can be adiabatically eliminated while mediating anti-$\mathcal{PT}$-symmetric dissipative coupling with two kinds of phase factors. Adjusting the relative phases allows the observation of \emph{periodic} Riemann-sheet structures with distributed exceptional points (Eps). At the Eps, destructive quantum interference breaks both the $\mathcal{T}$- and $\mathcal{P}$-inversion symmetry, resulting in unidirectional and chiral photon transmissions. In the reversed-dissipation regime, the nonreciprocal bandwidth is no longer limited by the mechanical mode linewidth but is improved to the linewidth of the cavity resonance. Furthermore, we find that the direction of the unidirectional and chiral energy transfer could be reversed by changing the \emph{parity} of the Eps. Extending non-Hermitian couplings to a three-cavity model, the broken anti-$\mathcal{PT}$-symmetry allows us to observe high-order Eps, at which a parity-dependent chiral circulator is demonstrated. The driving-phase controlled periodical Riemann sheets allow observation of the parity-dependent unidirectional and chiral energy transfer and thus provide a useful cell for building up nonreciprocal array and realizing topological, e.g., isolators, circulators, or amplifiers.
\end{abstract}




\end{frontmatter}


\section{Introduction}
\label{sec:introduction}
Nonreciprocal elements play an important role in both classical and quantum information processing~\cite{caloz2018electromagnetic,huang2018nonreciprocal}. Such devices can reduce backscattering induced interference and then protect the signal transmission~\cite{jalas2013and,metelmann2015nonreciprocal,sliwa2015reconfigurable,kerckhoff2015chip}. In principle, non-reciprocity can be achieved by breaking the time-reversal symmetry. Conventional nonreciprocal devices are mainly based on the magneto-optical responses in ferrite medium~\cite{stadler2013integrated}. However, the required large magnetic field is deleterious for many quantum information devices~\cite{you2007low,koch2007charge,blais2020quantum}. In recent years, there has been an active search for alternatives to magneto-optical non-reciprocity. Several magnetic-free strategies have been developed based on the, e.g., refractive index modulation~\cite{yu2009complete, lira2012electrically, sounas2017non}, $\mathcal{PT}$-symmetric structures~\cite{el2018non,regensburger2012parity,lin2011unidirectional, Ramezani2010Unidirectional,quijandria2018pt,zhang2015giant}, and out-of-phase temporal modulations~\cite{kamal2011noiseless,estep2014magnetic}.

Among various types of resonator-based quantum systems~\cite{gu2017microwave,kurizki2015quantum,krantz2019quantum,huang2020superconducting,xiang2013hybrid,clerk2020hybrid}, cavity optomechanical devices are becoming a versatile platform for on-chip controlling photon transfer~\cite{aspelmeyercavity2014,wang2014optomechanical,xiong2015review,liu2017electromagnetically,dong2012optomechanical,dong2015optomechanical,liu2018cavity,Wang:19,liao2016macroscopic}. Optomechanical systems have also been proposed as powerful alternatives to build nonreciprocal elements such as  isolators~\cite{hafezi2012optomechanically,miri2017optical,zhang2018optomechanical}, circulators~\cite{xu2015optical,barzanjeh2017mechanical,shen2018reconfigurable} and directional amplifiers~\cite{malz2018quantum,jiang2018directional,wanjura2020topological} for optics or microwaves. In optomechanical devices, the required broken time-reversal symmetry for non-reciprocity has been realized using, e.g., a synthetic gauge field~\cite{fang2017generalized,ruesink2016nonreciprocity,ruesink2018optical}, Brillouin momentum biases~\cite{shen2016experimental,kim2015non,kittlaus2018non,dong2015brillouin}, a spinning induced Sagnac frequency shifts~\cite{jiao2020nonreciprocal}, or Floquet frequency components~\cite{de2020nonreciprocal}. Breaking the time-reversal symmetry in cavity optomechanical devices is particularly appealing because strong non-reciprocity can be produced without needing a magnetic field or a nonlinear medium~\cite{verhagen2017optomechanical}. Depending on the achievements of the ground-state cooling of the mechanical resonators, nonreciprocal photon transmission and amplifications have been demonstrated in superconducting electromechanical devices with quantum-limited add-noise~\cite{Peterson2017demonstarte,bernier2017nonreciprocal,de2019realization,bernier2018nonreciprocity}. However, the bandwidth for nonreciprocal photon transmission is arguably limited to the Kilohertz to Megahertz (kHz-MHz) range for optomechanical devices~\cite{fang2017generalized,ruesink2016nonreciprocity,ruesink2018optical,shen2016experimental,kim2015non,kittlaus2018non,dong2015brillouin} or the Hertzthe to the Kilohertz (Hz-kHz) range for electro-mechanical devices~\cite{de2020nonreciprocal,verhagen2017optomechanical,Peterson2017demonstarte,bernier2017nonreciprocal,de2019realization,bernier2018nonreciprocity}. In the above approaches, the bandwidth limit is set by the linewidth of the mechanical oscillator resonance~\cite{xu2016nonreciprocal,sohn2018time,xu2019nonreciprocal}, which is usually much smaller than the corresponding cavity mode linewidth in the microwave or optical domain.

Recently, experimental studies of non-Hermitian optomechanics have shown that the mechanical linewidth can be close to or even much larger than the cavity linewidth~\cite{toth2017dissipative,Shen2021dissipative,toth2018maser,ohta2021rare}. When the optomechanical system enters the \textit{reversed-dissipation regime} where the mechanical decay rate is much larger than the cavity linewidth~\cite{nunnenkamp2014quantum}, the mechanical mode serves as a dissipative quantum reservoir and the dynamical backaction to the optical modes can no longer be ignored~\cite{lee2018quantum}. In such a reversed-dissipation regime, it has been discussed that the mechanical reservoir can introduce a new type of parametric instability~\cite{shomroni2019two} or be further developed to enhance optomechanical entanglement~\cite{zhang2020enhanced}.

In this work, we propose broadband optomechanical nonreciprocity in the reversed-dissipation regime, in which the nonreciprocal bandwidth is no longer limited by the mechanical linewidth. Reversing the dissipations of optical and mechanical modes allows the bandwidth to be dictated by the cavity linewidth. To be more specific, we study a multimode optomechanical system where two cavity modes optomechanically coupled to a common mechanical mode. In the reversed-dissipation regime, the mechanical mode with a large decay rate is adiabatically eliminated and then mediates a dissipative coupling between these two optical modes. Without loss of generality, the coherent coupling between cavity modes is also taken into consideration. By controlling the ratio between the two parameters, relative phase and coupling rate, the system energy levels exhibit a periodic Riemann sheet with distributed odd and even Eps. The reversed dissipation will not only overpass the nonreciprocal bandwidth, which is limited by the narrow mechanical linewidth presented in previous works~\cite{hafezi2012optomechanically,miri2017optical,zhang2018optomechanical,xu2015optical,barzanjeh2017mechanical,
shen2018reconfigurable,malz2018quantum,jiang2018directional,wanjura2020topological,ruesink2016nonreciprocity,fang2017generalized,
ruesink2018optical,shen2016experimental,kim2015non,kittlaus2018non,dong2015brillouin,de2020nonreciprocal,jiao2020nonreciprocal,
verhagen2017optomechanical,Peterson2017demonstarte,bernier2017nonreciprocal,de2019realization,bernier2018nonreciprocity,
xu2016nonreciprocal,sohn2018time,xu2019nonreciprocal} but also result in a broken $\mathcal{P}$-inversion symmetry and chiral energy transfer. In the following, we present a \textit{parity-dependent} unidirectional and chiral energy transfer, the direction of which could be totally reversed by toggling the parity of Eps.

\section{The model}
\begin{figure}[t]\centering
\includegraphics[scale=0.42]{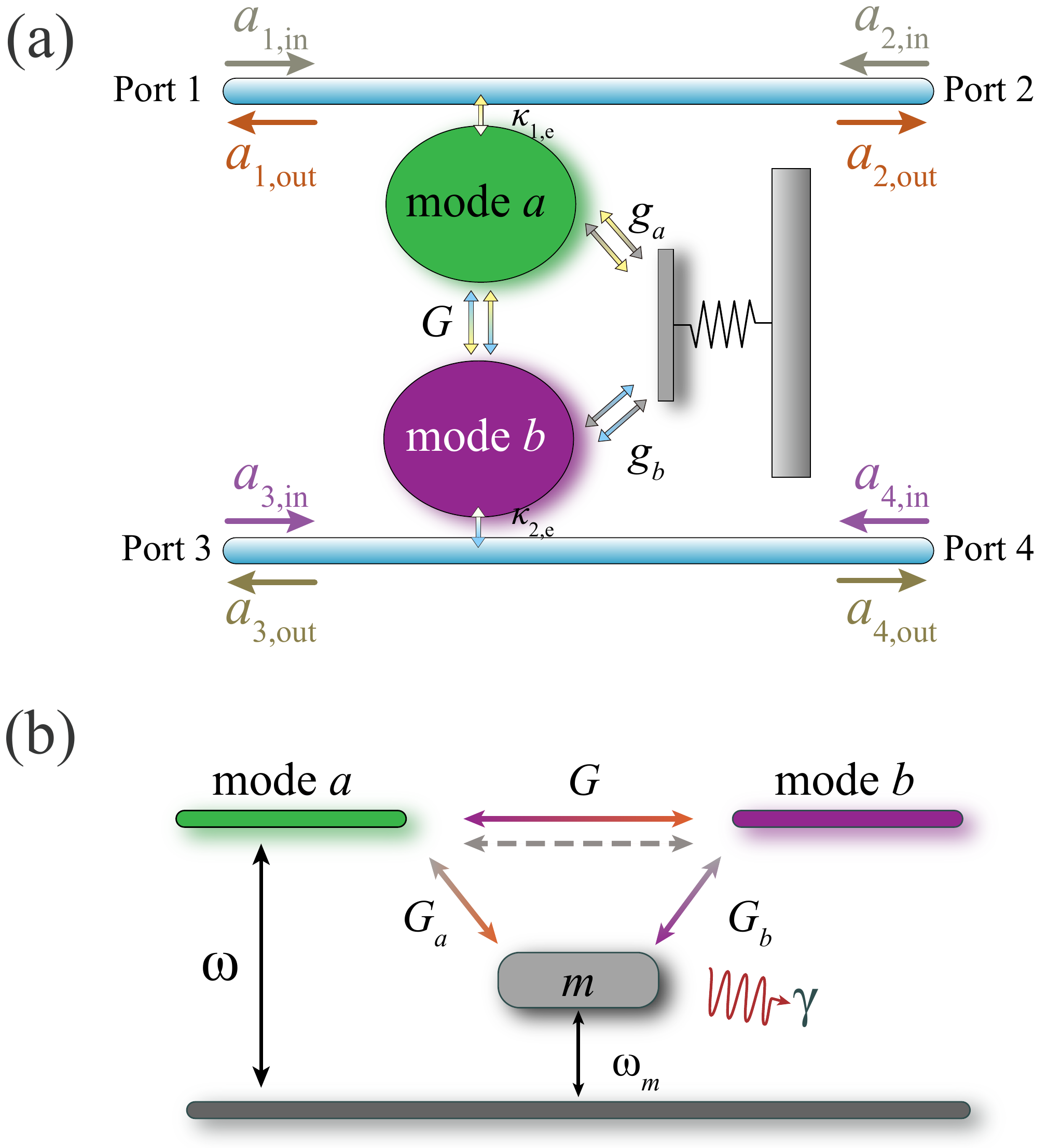}
\caption{\textbf{(a)}, Schematic illustration of the cavity optomechanical system. $a_{\mathrm{j,in}}$ ($a_{\mathrm{j,out}}$) denotes the input (output) signal amplitude from the j-th port. $\kappa_{\mathrm{1,e}}$ ($\kappa_{\mathrm{2,e}}$) denotes the external loss of mode $a$ ($b$), respectively. The cavity mode $a$ ($b$) optomechanically couples to the mechanical mode $m$, and $g_{a}$ ($g_{b}$) is the single-photon coupling strength, which can be greatly enhanced and linearized with an external strong pump field. \textbf{(b)}, Energy level diagram of the system. The linearized effective optomechanical coupling strengths are given by $G_a$ and $G_b$, respectively. }\label{fig1}
\end{figure}
As shown in Fig.~\ref{fig1} (a), our model consists of two bosonic cavity modes and a low-Q mechanical resonator. Two cavity modes couple to each other with a coupling strength $G$, and both of them optomechanically couple to the mechanical mode $m$ with the single-photon coupling strength given by $g_{a}$ and $g_{b}$, respectively. The above system can be well described by the following non-Hermitian Hamiltonian:
\begin{eqnarray}
\label{Hnon}
H  &=&\left( \omega_a -i\kappa_{1}\right) a^{\dag}a+\left(\omega_b -i\kappa_{2}\right) b^{\dag}b\\ \nonumber
&+&\left( \omega_m -i\gamma\right) m^{\dag}m +G( a^{\dag }b+b^{\dag }a)\\  \nonumber
&+&g_{a}a^{\dag}a(m+m^{\dag})+g_{b}b^{\dag}b(m+m^{\dag}),
\end{eqnarray}

\noindent here we set $\hbar=1$. $a (a^\dagger)$, $b (b^\dagger)$ and $m (m^\dagger)$ represent the bosonic annihilation (creation) operators with resonant frequencies $\omega_a$, $\omega_b$ and $\omega_m$, respectively. $\kappa_{1}=\kappa_{1,i}+\kappa_{1,e}$ ($\kappa_{2}=\kappa_{2,i}+\kappa_{2,e}$) denotes the loss for cavity modes $a$ ($b$), where $\kappa_{1,i}$ ($\kappa_{1,e}$), $\kappa_{2,i}$ ($\kappa_{2,e}$) is the internal, external loss of mode $a$ ($b$), respectively. $\gamma$ represents the dissipation of the mechanical mode $m$.

To linearize the Hamiltonian~(\ref{Hnon}), we can introduce two driving tones to the system: $i(\varepsilon_{1}a^{\dagger}e^{-i\Omega t}e-H.c.)+i(\varepsilon_{2}b^{\dagger}e^{-i\Omega t}-H.c.)$, then the linearized Hamiltonian reads
\begin{eqnarray}
\label{Hnon1}
\tilde{H}  &=&\left( \Delta_a -i\kappa_{1}\right) a^{\dag}a+\left(\Delta_b -i\kappa_{2}\right) b^{\dag}b\\ \nonumber
&+&\left( \omega_m -i\gamma\right) m^{\dag}m +G( a^{\dag }b+b^{\dag }a)\\ \nonumber
&+&G_{a}a^{\dag }m+G_{a}^{*}m^{\dag }a+G_{b} b^{\dag }m+G_{b}^{*}m^{\dag }b.
\end{eqnarray}

\noindent $\Delta_a=\omega_a-\Omega~(\Delta_b=\omega_b-\Omega)$ is the frequency detuning between driving tones and the mode $a$ ($b$). $G_{a} (G_{b})$ is the linearized optomechanical coupling strength between mode $a (b)$ and mechanical mode $m$. The energy level diagram of the linearized system is presented in Fig.~\ref{fig1}(b). In the reversed-dissipation regime, the decay rate of the mechanical mode is much larger than both the decay rates of the cavity modes and the optomechanical couplings strengths, viz., $\gamma\gg (G_{a},G_{b}, \kappa_{1}, \kappa_{2})$. After adiabatically eliminating the mechanical mode $m$, a strong dark state interaction takes place (see Supplementary Materials section 1 for details), and the effective Hamiltonian for the whole system can be expressed as
 \begin{eqnarray}
 \label{Hnon2}
H_{\rm_{eff}}  &=&\left( \omega -i\kappa \right) a^{\dag}a+\left( \omega -i\kappa \right) b^{\dag }b\\ \nonumber
&+&\left(iJ\mathrm{e}^{-i\theta} +G\right) a^{\dag }b+\left(iJ\mathrm{e}^{i\theta} +G\right) b^{\dag }a.
\end{eqnarray}

\noindent Here, we assume $|G_a|=|G_b|$ and $J=|G_a|^2/\gamma$ is the effective dissipative coupling strength. In a reference frame rotating at the driving frequency $\Omega$, the modes $a$, $b$ and $m$ could be assumed to be degenerate (i.e., $\Delta_a=\Delta_b=\omega_m =\omega$), and the cavity mode $a$, $b$ has the same dissipation rate $\kappa$.

The eliminated mechanical mode $m$ now has two impacts on the Hamiltonian~(\ref{Hnon2}). First, it produces dissipative coupling terms $iJ(\mathrm{e}^{-i\theta}a^{\dag }b+ \mathrm{e}^{i\theta}b^{\dag }a)$ to the system. Notably, dissipative coupling is recognized as anti-$\mathcal{PT}$ symmetric and has become an important concept of non-Hermitian physics in recent years~\cite{zhao2020observation,jiang2021energy,yu2019prediction,PhysRevA.104.012218,PhysRevB.103.235110,PhysRevB.102.161101}. Correspondingly, the coherent coupling terms $G(a^{\dag}b+ b^{\dag}a)$ are $\mathcal{PT}$ symmetric (see Supplementary Materials section 2 for details). Another aspect is that it increases the total losses of modes $a$ and $b$, that is, $\kappa=\kappa_{\mathrm{i}}+\kappa_{\mathrm{e}}+J$. In the following discussion, without loss of generality, we assume that modes $a$ and $b$ are always under the critical-coupling condition , i.e., $\kappa=2\left(J+\kappa_{\mathrm{i}}\right)$.

Considering both the coherent and dissipative couplings for cavity modes $a$ and $b$, the total interaction Hamiltonian is then shown with the detailed expression as follows:
\begin{equation}
H_{\mathrm{I}} =\left( G+iJ\mathrm{e}^{-i\theta}\right) a^{\dag }b+\left( G+iJ\mathrm{e}^{i\theta}\right) b^{\dag }a.
\label{Htdetail}
\end{equation}
Notably, the mechanical mode mediates two kinds of phase factors in dissipative coupling. One mediated phase factor is fixed at $\frac{\pi}{2}$, which is represented by the imaginary unit in Eq.~(\ref{Htdetail}). The other phase factor is determined by the phase difference between two parametric optomechanical couplings, i.e., $\theta={\rm{arg}}(G_{a}^{*}G_b)$. In the experiment, such a parametric phase is \emph{in-situ} tunable by controlling the phases between the external pump tones. The above Eq.~(\ref{Htdetail}) indicates that destructive quantum interference occurs between coherent and dissipative couplings at a special phase factor. It is also the key factor to the unidirectional and chiral properties discussed in this work.

\section{Phase controlled periodical Riemann-sheets and odd-even exceptional points}
\begin{figure*}[t]
\includegraphics[scale=0.35]{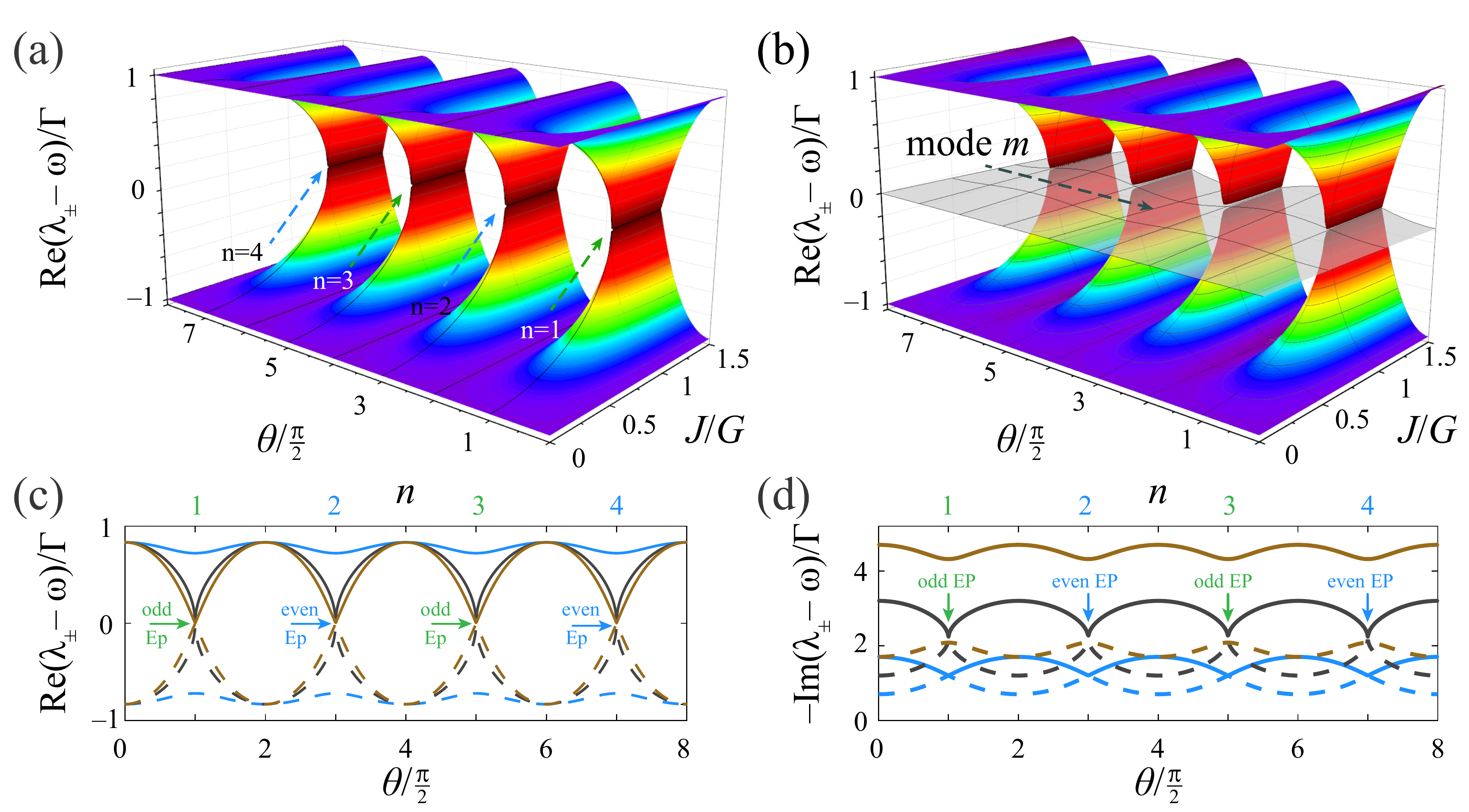}\centering
\caption{Riemann sheet structure of the eigenvalues and eigenvalues evolution. \textbf{(a)}, Supermode frequencies (real parts of $\lambda_{\pm}$) of Eq~(\ref{eigens}) versus phase factor $\theta$ and the dissipative coupling strength $J$. Here, the coherent coupling strength is fixed at $G/\kappa_{\mathrm{i}}$=10 for the calculation. The cavity internal decay rate can be $\kappa_{i}=$~13~kHz for microwave cavity~\cite{Peterson2017demonstarte}, or $\kappa_{i}=1$~MHz for optical cavity~\cite{kim2015non}. \textbf{(b)}, Eigenvalues of Hamiltonian~(\ref{Hnon1}). The dissipation rate of the mechanical mode is set to $\gamma/G=50$, and the other parameters are the same with (a). \textbf{(c)}, \textbf{(d)}, The real parts and imaginary parts of Eq~(\ref{eigens}) versus phase factor $\theta$ with $J/G=0.5$ $(\rm{blue})$, 1 $(\rm{black})$, 1.5 $(\rm{brown})$, respectively. The parameters used are the same as in (a). The green (blue) arrow marks the odd (even) Eps.}
\label{fig2}
\end{figure*}

To reveal the odd-even exceptional points, we start by studying how the parametric phase $\theta$ affects the energy level evolution. The corresponding complex eigenvalues of $H_{\rm_{eff}}$ are given as
\begin{equation}
\label{eigens}
\lambda_{\pm }=\omega -i\kappa \pm \sqrt{\left( iJ\mathrm{e}^{-i\theta}+G\right) \left( iJ\mathrm{e}^{i\theta}+G \right)}.
\end{equation}
The eigenfrequencies $\omega_{\pm}$ and the corresponding linewidthes are the real and imaginary parts of $\lambda_{\pm}$, respectively. To verify the validity of the adiabatical approximation, we also perform numerical calculations of eigenfrequencies of Hamiltonian~(\ref{Hnon1}), as shown in Fig~\ref{fig2} (b). Compared to Fig.~\ref{fig2} (a), Fig~\ref{fig2} (b) also shows a similar Riemann sheets structure with extra eigenvalues of mode $m$ [semitransparent surface in Fig~\ref{fig2} (b)]. However, in the reversed-dissipation regime, the mechanical mode $m$ is almost decoupled to the Riemann sheet structure. Thus, the adiabatical elimination of mode $m$ is appropriate in our analyses.

To more clearly reveal the periodic coincidence and repulsion of eigenvalues, we plot curves of the eigenfrequencies and linewidths for $J/G=(0.5,1,1.5)$ in Fig.~\ref{fig2} (c) and~\ref{fig2} (d). Indeed, the non-Hermitian degeneracy and exceptional points (Eps) exist only when the dissipative and coherent couplings are balanced, viz., $J=G$. When the dissipative and coherent couplings are not balanced, either frequencies ($J<G$) or linewidths ($J>G$) will not coincide.

As shown in Figs.~\ref{fig2}, the eigenfrequencies exhibit \emph{periodic} a Riemann-sheet structure as a function of phase $\theta$ and the ratio between dissipative and coherent coupling strength $J/G$. The eigenfrequencies and the corresponding linewidths merge into the Eps under certain phase-matching and balanced dissipative-coherent coupling strength conditions. Notably, the reversed-dissipation mechanical mode implants a fixed phase factor at $\frac{\pi}{2}$ and a tunable parametric phase $\theta$ into the dissipative coupling terms. These Eps require the ratio between parametric and fixed phases to satisfy the phase-matching condition $\theta/\frac{\pi}{2} = 2n-1$, $n\in \mathbb{Z}$.

Each integer $n$ specifies a group of Riemann surfaces accompanied by one exceptional point. Odd (even) $n$ specifies odd (even) Eps, respectively. The Eps have been widely explored in coupled two-component systems, but usually only show a single exceptional point~\cite{PhysRevA.95.013843, PhysRevA.96.023812, zhang2021exceptional, zhao2020observation,jiang2021energy,yu2019prediction,PhysRevA.104.012218,PhysRevB.103.235110,PhysRevB.102.161101, doppler2016dynamically,zhang2019dynamically,xu2016topological,PhysRevA.104.063508}. In this work, as shown in Fig.~\ref{fig2}, periodical Riemann sheets and odd-even Eps occur when the phase-matching condition is satisfied. In addition to the reported schemes such as controlling the frequency detuning~\cite{zhang2017observation} and the decay rates~\cite{peng2014loss}, we propose the use of an in-situ tunable phase $\theta$ to observe the Riemann sheet.

The transmission spectrum is a powerful tool to observe energy level evolution, as well as the emergence of Eps. Considering a probe signal (with frequency $\omega_{p}$) input from port 1, the quantum Langevin equations can be expressed as,
\begin{eqnarray}
\dot{a} &=&-\left( i\delta +\kappa \right) a+\left( J\mathrm{e}^{-i\theta}-iG\right) b+\sqrt{\Gamma }a_{\mathrm{in}}, \nonumber\\
\dot{b} &=&-\left( i\delta +\kappa \right) b+\left( J\mathrm{e}^{i\theta}-iG\right) a.
\label{eqlang}
\end{eqnarray}
The steady-state solutions of Eqs.~(\ref{eqlang}) are,
\begin{eqnarray}
\left\langle a\right\rangle  &=&\frac{\left( i\delta +\kappa \right) \sqrt{\Gamma }a_{\mathrm{1,in}}}{\left( i\delta +\kappa \right) ^{2}-\left(J\mathrm{e}^{-i\theta}-iG\right) \left( J\mathrm{e}^{i\theta}-iG\right) }, \nonumber\\
\left\langle b\right\rangle  &=&\frac{\left( J\mathrm{e}^{i\theta}-iG\right) \sqrt{\kappa }a_{\mathrm{1,in}}}{\left( i\delta +\kappa \right)^{2}-\left(J\mathrm{e}^{-i\theta}-iG\right) \left(J\mathrm{e}^{i\theta}-iG\right) }.
\end{eqnarray}
Using the input-output relationship
\begin{equation}
a_{\mathrm{2,out}}=a_{\mathrm{1,in}}-\sqrt{\kappa}\left\langle a\right\rangle.
\end{equation}
The transmission spectrum for energy transfer from port 1 to 2 is given by
\begin{equation}
S_{21}=\frac{a_{\mathrm{2,out}}}{a_{\mathrm{1,in}}}=1-\frac{\kappa \left(i\delta+\kappa \right) }{\left( i\delta +\kappa \right) ^{2}-\left( J\mathrm{e}^{-i\theta}-iG\right) \left( J\mathrm{e}^{i\theta}-iG\right) },
\end{equation}
where $\delta=\omega-\omega_{p}$ is the frequency detuning between cavity mode resonance $\omega$ and frequency $\omega_{p}$ of the input probe signal.

\begin{figure}[t]
\includegraphics[scale=0.35]{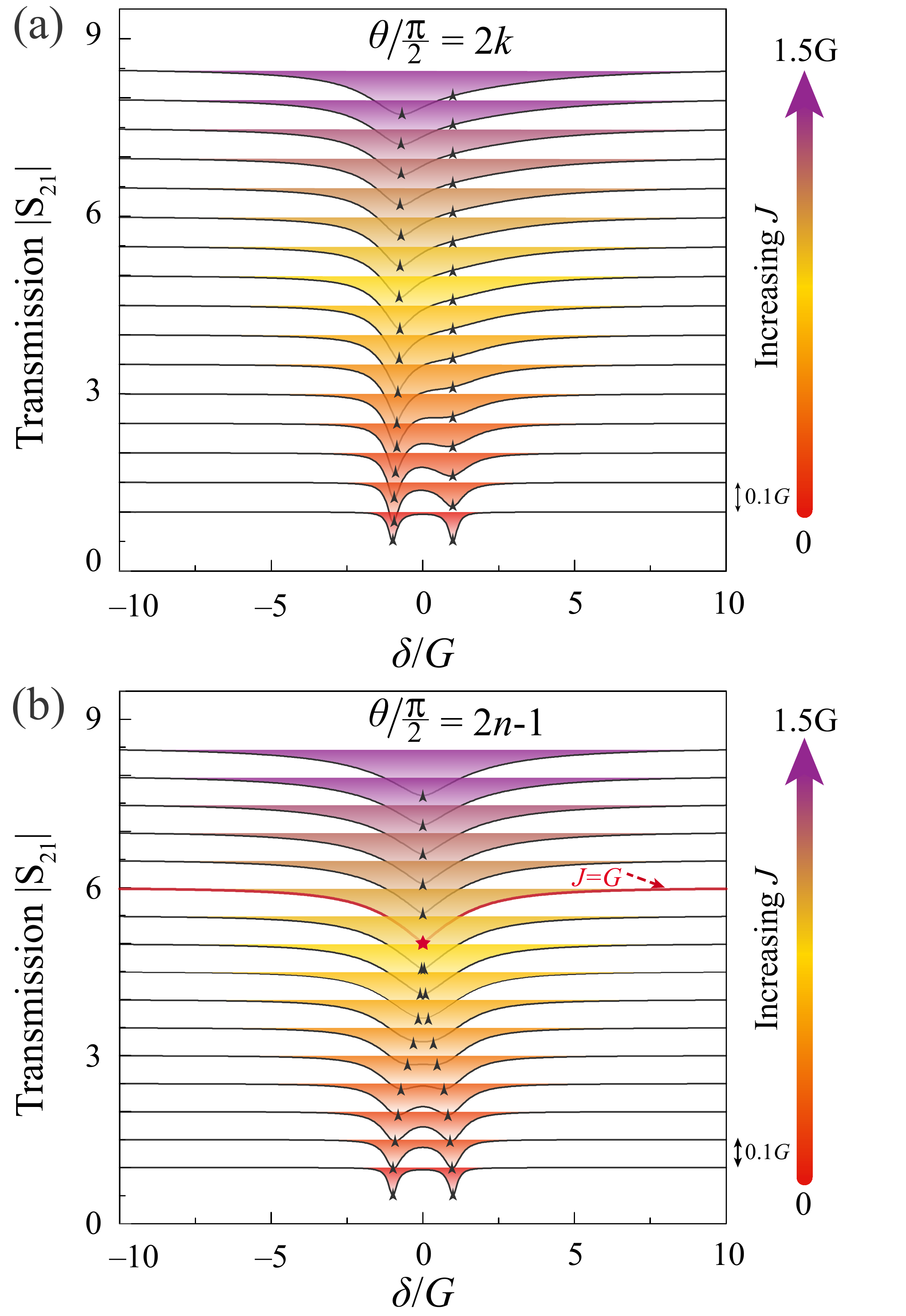}\centering
\caption{The transmission spectrum $S_{21}$ is plotted for various dissipative coupling strengths. The phase is fixed at $\theta/\frac{\pi}{2} =2k, k\in \mathbb{Z}$ for \textbf{(a)}, and $\theta/\frac{\pi}{2} =2n-1, n\in \mathbb{Z}$ for \textbf{(b)}, respectively. For each figure, from bottom to top (labelled by the arrow bar), the dissipative coupling strength increased from 0 to $1.5~G$ with a step $0.1~G$. The red pentagram inside (b) marks Ep with phase-matching and balanced dispersive and dissipative coupling strengths. The values for other parameters are the same as those used in Fig.~\ref{fig2}.}\label{fig3}
\end{figure}

Figure~\ref{fig3} shows the transmission spectrum for a variety of dissipative couplings strengths. When the phase matching conditions are not satisfied, e.g., setting $\theta/\frac{\pi}{2}=2k$, $k\in \mathbb{Z}$ for the plots in Fig.~\ref{fig3} (a), the eigenvalues are now reduced to $\lambda_{\pm ,\theta =k\pi}=\omega \pm G-i\left(\kappa \pm J\right)$. In this case, the coherent ($G$) and dissipative coupling ($J$) strengths only affect the supermode eigenfrequency and linewidth, respectively. As shown in Fig.~\ref{fig3} (a), the eigenfrequency positions remain unchanged. However, the linewidths become wider with continuously increasing dissipative coupling strength, resulting in an asymmetric supermode readout.

As discussed above, non-Hermitian degeneracy occurs and the Eps can be observed when the phase-matching conditions are satisfied, viz., $\theta/\frac{\pi}{2}=2n-1$, $n\in \mathbb{Z}$. Actually, as shown in Fig.~\ref{fig3} (b), the energy-level attracts and further collapses into the exceptional points as the dissipative coupling increases until it equilibrates with the coherent coupling. Notably, the evolution of the energy spectrum does not depend on the parity of Eps.

\section{Parity dependent unidirectional photon transport}

Apart from energy-level attraction, we demonstrate that the Hamiltonian becomes naturally unidirectional with cascaded-like interaction terms at these Eps. We show in the latter that the parity of Eps will greatly affect the direction of the unidirectional and chiral photon transport behavior. In particular, at odd Eps (i.e., when $n$ is an odd number), the Hamiltonian in Eq.~(\ref{Hnon2}) becomes
\begin{equation}
\tilde{H}_{\mathrm{odd}} =\left( \omega-i\kappa \right) \left( a^{\dag }a+b^{\dag}b\right) +2Ga^{\dag }b.  \label{ODD}
\end{equation}
However, for even Eps (i.e., when $n$ is an even number), the Hamiltonian becomes
\begin{equation}
\tilde{H}_{\mathrm{even}} =\left( \omega-i\kappa \right) \left( a^{\dag }a+b^{\dag}b\right) +2Gb^{\dag }a.  \label{EVEN}
\end{equation}
Equations~(\ref{ODD}) and (\ref{EVEN}) are $\mathcal{T}$ ($\mathcal{P}$)-inversion asymmetry. Thus, energy transfers from mode $a$ to $b$ should be unidirectional. In particular, the direction of nonreciprocal photon transport depends on the parity of the phase-match factor $n$.
\begin{figure}[t]
\includegraphics[scale=0.38]{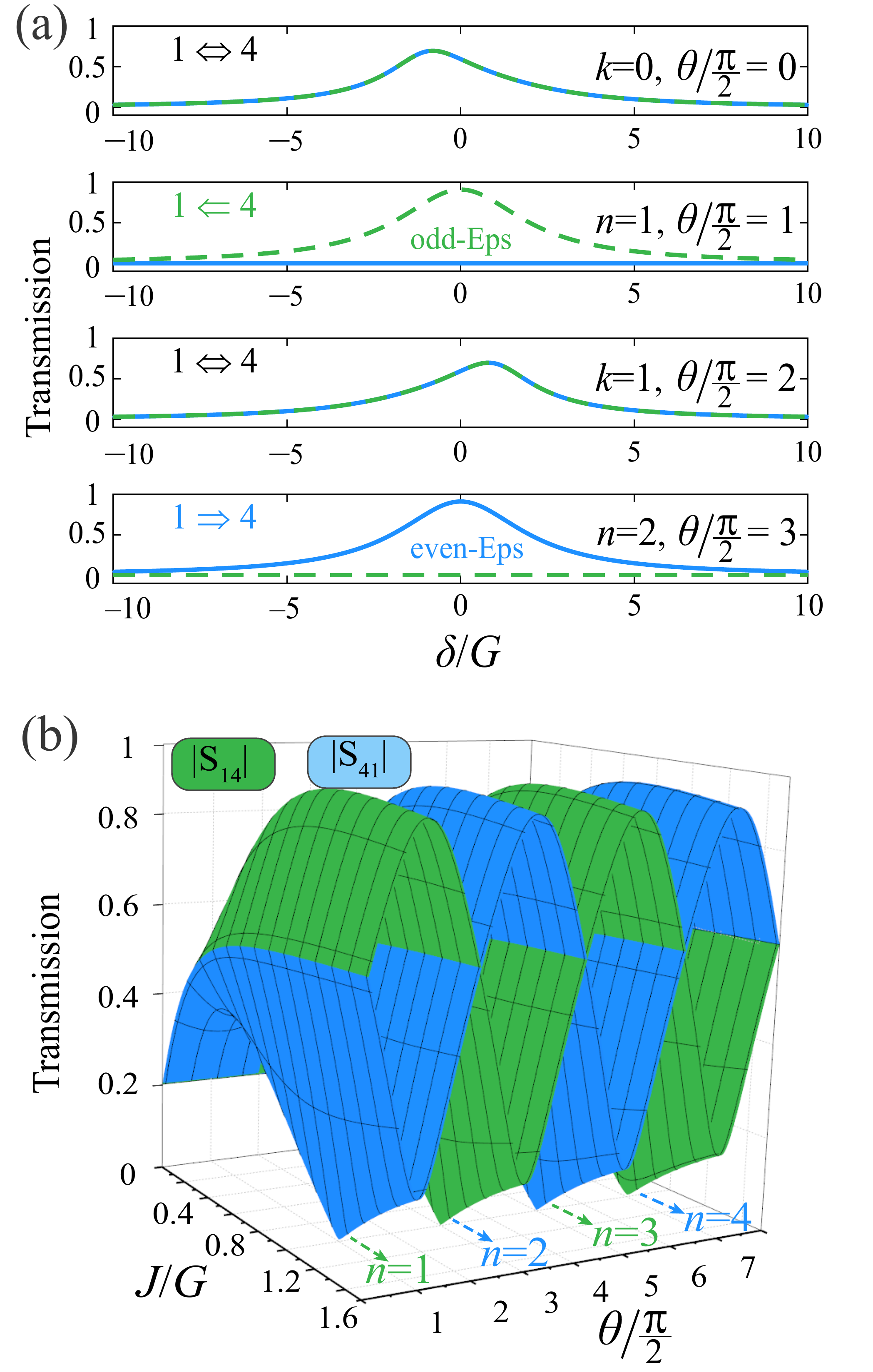}\centering
\caption{Nonreciprocal photon transmission. \textbf{(a)}, Transmission spectrum from port i to j defined as $|S_{ji}|=|a_{\mathrm{j,out}}/a_{\mathrm{i,in}}|$ are plotted versus probe detunings $\delta$ for $\theta/\frac{\pi}{2}=(0,1,2,3)$ with $J=G$. Blue-solid (green-dashed) curves represent the transmission $|S_{41}|$ ($|S_{14}|$), respectively. \textbf{(b)}, 3D plot of the transmission coefficients $|S_{41}|$ and $|S_{14}|$ versus $J/G$ and phase $\theta$ to show an ideal nonreciprocity (unidirection) at the odd or even Eps.}\label{fig4}
\end{figure}

To demonstrate how the parity of the Eps affects the photon transport properties, we take ports 1 and 4 as an example to study its transmission spectrum. The corresponding transmission coefficient is
\begin{equation}
S_{41}= \frac{a_{\mathrm{4,out}}}{a_{\mathrm{1,in}}} = \frac{\kappa \left( J\mathrm{e}^{i\theta}-iG\right) }{\left( i\delta +\kappa \right) ^{2}-\left( J\mathrm{e}^{-i\theta}-iG\right) \left( J\mathrm{e}^{i\theta}-iG\right) } ,\label{uni1to4}
\end{equation}
and the reversed photon transmission (from port 4 to 1) is given by
\begin{equation}
S_{14}= \frac{a_{\mathrm{1,out}}}{a_{\mathrm{4,in}}} = \frac{\kappa \left(J\mathrm{e}^{-i\theta}-iG\right) }{\left( i\delta +\kappa \right) ^{2}-\left( J\mathrm{e}^{i\theta}-iG\right) \left( J\mathrm{e}^{-i\theta}-iG \right) }.\label{uni4to1}
\end{equation}

Figure.~\ref{fig4} (a) shows the transmission spectrum between ports 1 and 4 for four typical phase-factors. When $\theta/\frac{\pi}{2}=2k, k\in\mathbb{Z}$ [as an example with $k=(0,1)$ and shown in Fig.~\ref{fig4} (a)], photons transfer evenly and reciprocally between ports 1 and 4. This can be physically understood by recalling the eigenvalues evolution from Eq.~(\ref{eigens}), which now are further simplified to $\lambda_{\pm ,\theta =k\pi}=\omega \pm G-i\left(\kappa \pm J\right)$. It is obvious that the coherent coupling shifts the eigenfrequency and the dissipative coupling alters the linewidth. As a result, destructive interference does not occur and the photons can evenly transport between ports 1 and 4. In contrast, as phase matching conditions are satisfied, viz., $\theta/\frac{\pi}{2}=2n-1, n\in\mathbb{Z}$, destructive interference occurs and the photon transfer between ports 1 and 4 becomes nonreciprocal. As an example with $n=(1,2)$ and shown in Fig.~\ref{fig4} (a), the photon transfer becomes totally unidirectional at the Eps.

It is notable that the direction for unidirectional photon transfers is now determined by the parity of the phase-match factor $n$. More specifically, when $n$ is odd, as an example $n=1$ corresponding to $\theta/\frac{\pi}{2}=1$, the photon transports between ports 1 and 4 are plotted and shown in Fig.~\ref{fig4} (a). It clearly shows that the photons transfer from port 4 to 1 ($4\Rightarrow1$) but are totally forbidden in turn ($1\nRightarrow4$). When $n$ is even, the exact opposite transport behaviors are presented. As an example $n=2$ corresponding to $\theta/\frac{\pi}{2}=3$, the unidirectional performance becomes inverted. The photons only transfer from port 1 to 4 ($1\Rightarrow4$) but are totally forbidden in turn ($4\nRightarrow1$).

In addition to the unidirectional transfer with balanced coherent and dissipative couplings, the more general photon transmission is presented in Fig.~\ref{fig4} (b) by plotting $|S_{14}|$, and $|S_{41}|$ versus both the ratio $J/G$ and phase-factor $\theta$. At Eps, photon transfer becomes ideally unidirectional. The direction is parity-dependent on $n$. Around these Eps, although the photon transfer is not unidirectional, it is still nonreciprocal. The observed unidirectional energy transfer at the Eps can be understood by the $\mathcal{T}$-inversion-symmetry broken that is induced by destructive quantum interference between coherent and dissipative mode couplings. For odd-$n$, the energy coupling term $b^{\dagger}a$ is totally eliminated, but the coupling rate for $a^{\dagger}b$ is enhanced by twice. For even-$n$, the coupling term $a^{\dagger}b$ is totally eliminated by the destructive interference. However, the coupling rate for $b^{\dagger}a$ is enhanced by twice. The parity of $n$ results in totally opposite interference cancellation between the couplings of modes $a$ and $b$, finally introducing parity-dependent unidirectional photon transfer.

To clearly show how the mechanical mode decay rate affects the photon unidirectional transmission bandwidth, we plot the nonreciprocal area $|S_{14}|$-$|S_{41}|$ at $\theta=\frac{\pi}{2}$ for various mechanical dissipation rates $\gamma$. Because the adiabatic approximation is not valid for the low mechanical dissipation case, in Fig.~\ref{fig5}, we use the full Hamiltonian method for the calculation (see Supplementary Materials section 3 for details). As shown in Fig.~\ref{fig5}, the nonreciprocal bandwidth increases with the mechanical-mode dissipation rate $\gamma$. In the reversed-dissipation regime, e.g., the purple-solid curve with $\gamma/G=10$, the nonreciprocal transmission bandwidth arrives at the level of the cavity linewidth. Upon further increasing the mechanical decay rate, e.g., the green-solid curve with $\gamma/G=50$, the nonreciprocal transmission bandwidth remains almost unchanged. Thus, in the reversed-dissipation regime, the bandwidth of the unidirectional photon transfer is identical to the decay rate $\kappa$. For previous optomechanical approaches~\cite{hafezi2012optomechanically,miri2017optical,zhang2018optomechanical,xu2015optical,barzanjeh2017mechanical,shen2018reconfigurable,malz2018quantum,jiang2018directional,wanjura2020topological,ruesink2016nonreciprocity,fang2017generalized,ruesink2018optical,shen2016experimental,kim2015non,kittlaus2018non,dong2015brillouin,de2020nonreciprocal,jiao2020nonreciprocal,verhagen2017optomechanical,Peterson2017demonstarte,bernier2017nonreciprocal,de2019realization,bernier2018nonreciprocity,xu2016nonreciprocal,sohn2018time,xu2019nonreciprocal}, because the mechanical mode decay rate is always much smaller than both the coupling rate and cavity decay rate, the nonreciprocal transmission bandwidth is limited to the low dissipation mechanical mode, as shown by the blue curve in Fig.~\ref{fig5}. In our scheme, the bandwidth of the unidirectional photon transfer at odd and even Eps can be improved to the level of the cavity decay rates, e.g., on the order of megahertz for electromechanics or tens of megahertz for optomechanics.
\begin{figure}[t]
\includegraphics[scale=0.6]{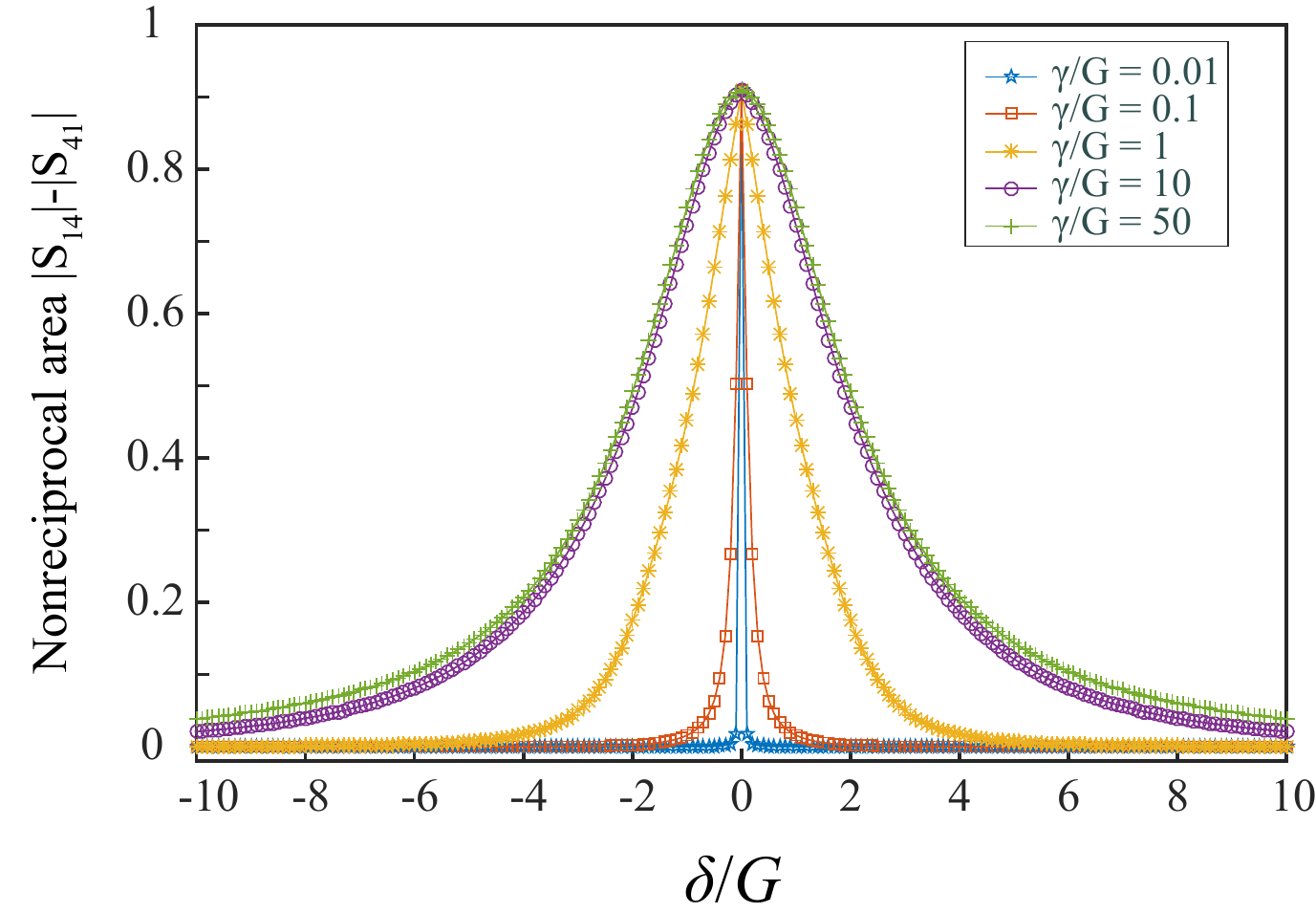}\centering
\caption{Nonreciprocal area as a function of probe detunings $\delta$. The parametric phase is fixed at $\theta/\frac{\pi}{2}=1$ with phase-matching factor $n$=1. The other parameters for calculation are the same as those used in Fig.~\ref{fig2}.}\label{fig5}
\end{figure}

\section{Chirality of the photon transport}

\begin{figure}[t]
\includegraphics[scale=0.4]{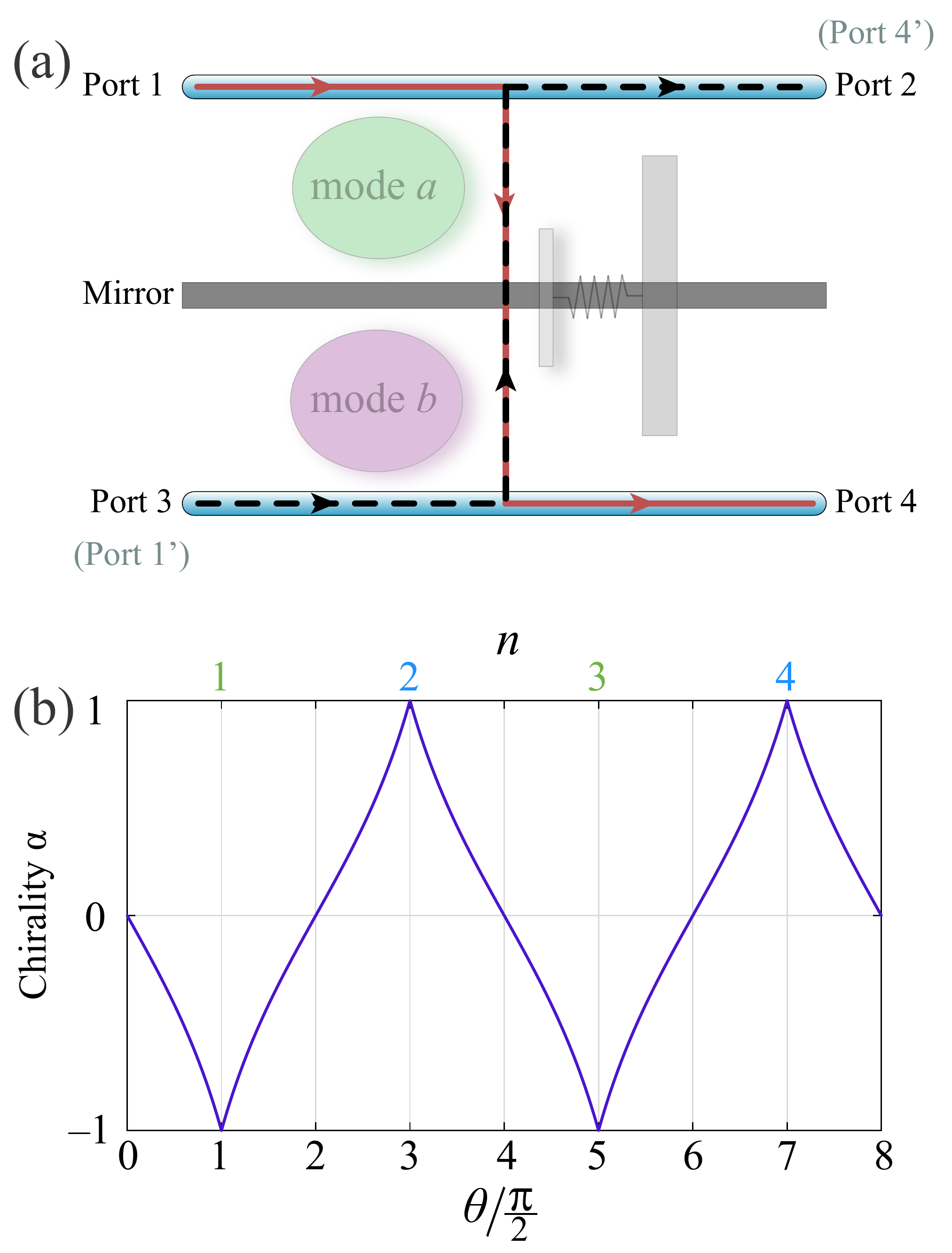}\centering
\caption{\textbf{(a)}, Schematic of the chirality of the system. Port 3 is the mirror image of port 1, while port 2 is the mirror image of port 4. \textbf{(b)}, Chirality as a function of phase $\theta$ (bottom x-axis) and the corresponding phase-matching number $n$ (top x-axis).}\label{fig6}
\end{figure}

In the previous section, we use ports 1 and 4 as an example to show the parity-dependent unidirectional photon transport at the Eps. Similar to mirrored transport, photon transmission between ports 3 and 2 can also be unidirectional at Eps but with totally opposite behaviors. The reversed mechanical mode dissipation will not only break the $\mathcal{T}$ inversion symmetry (introducing nonreciprocal transmission as discussed above) but could also result in a broken $\mathcal{P}$ inversion symmetry. Thus, the chiral energy transfer could be accessible. As shown in Fig.~\ref{fig6} (a), the energy transfers from port 1 to 4 can be defined as clockwise $\circlearrowright$ transfers. As a mirror image, energy transfer from ports 3 to 2 can be viewed as counterclockwise $\circlearrowleft$ transfers. For odd $n$, counterclockwise $\circlearrowleft$ light transfers are allowed, while clockwise $\circlearrowright$ photon transfers are forbidden. However, for even $n$, the situation becomes opposite, i.e.  $\circlearrowright$ is allowed, while $\circlearrowleft$ is forbidden. We can define a chirality as $\alpha=(|S_{41}|-|S_{23}|)/(|S_{41}|+|S_{23}|)$. As shown in Fig.~\ref{fig6} (b), the chirality $\alpha$ is plotted as a function of phase $\theta$ and its corresponding phase-matching number $n$. The extreme chirality $\alpha=\pm1$ depends on the parity of integer $n$.

\section{Parity-dependent chiral circulator at the high-order exceptional points}
\begin{figure}[t]
\includegraphics[scale=0.42]{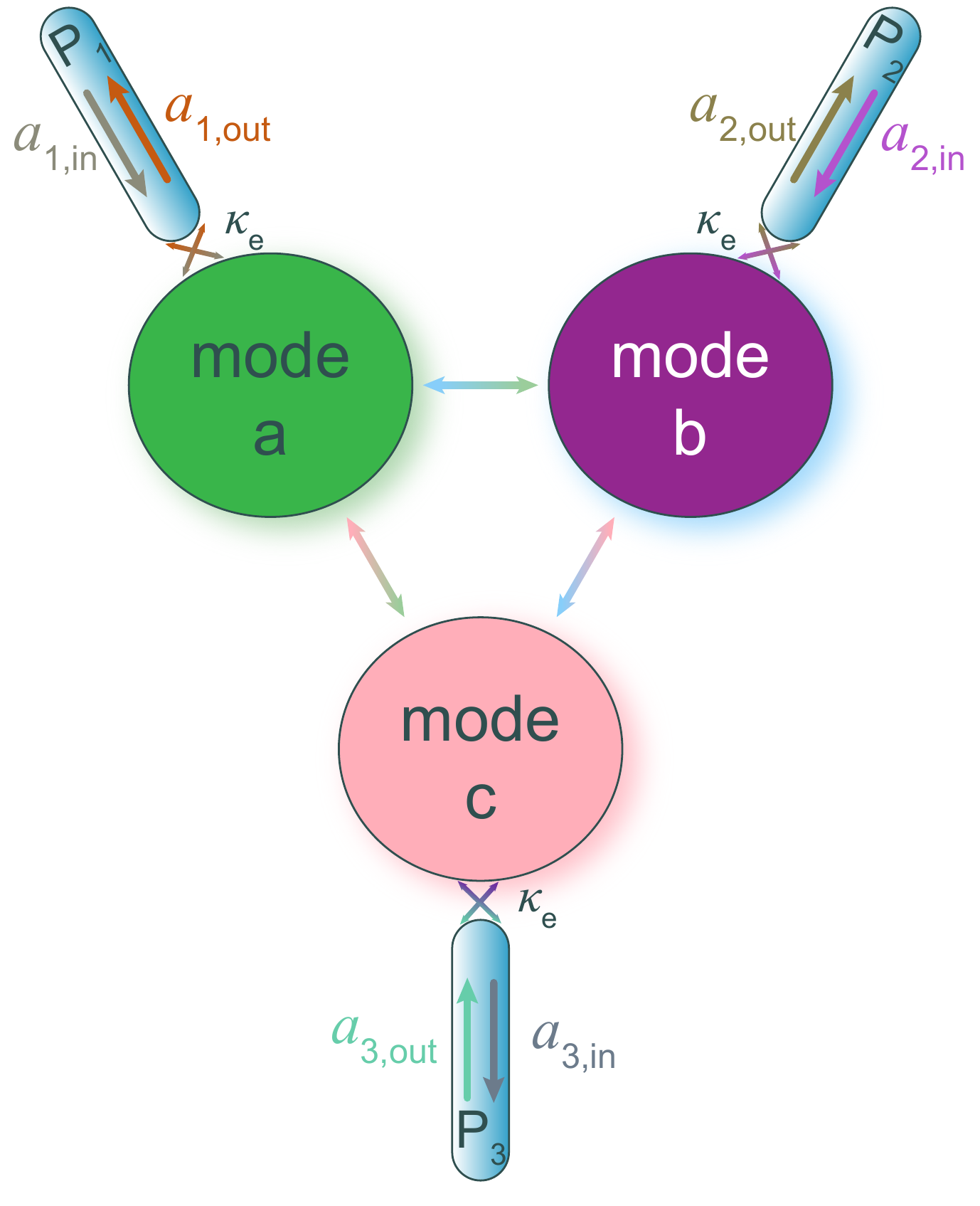}\centering
\caption{Scheme diagram for three mutual coupled cavity resonators. Every two cavities are coherently and dissipatively coupled simultaneously (marked by double-arrow) with identical coupling strengths.}\label{fig7}
\end{figure}
We now extend our scheme to the three-mode case, as shown in Fig.~\ref{fig7}, where every two modes are coupled with both coherent and dissipative couplings. For the three-mode case, the Hamiltonian is,
\begin{eqnarray}
H &=&\left( \omega -i\kappa \right) \left(a^{\dag }a+b^{\dag }b+ c^{\dag }c\right) \nonumber \\
&&+\left( iJ\mathrm{e}^{-i\theta}+G\right) a^{\dag}b+\left(iJ\mathrm{e}^{i\theta}+G\right) b^{\dag}a \nonumber\\
&&+\left( iJ\mathrm{e}^{-i\theta}+G\right) b^{\dag }c+\left( iJ\mathrm{e}^{i\theta}+G\right) c^{\dag}b \nonumber\\
&&+\left( iJ\mathrm{e}^{-i\theta}+G\right) c^{\dag }a+\left( iJ\mathrm{e}^{i\theta}+G\right) a^{\dag}c.  \label{Hthree}
\end{eqnarray}
Similar to the method in Supplementary Materials section 3, the corresponding Langevin equations can be expressed as,
\begin{equation}
\dot{\mu}=-iM\mu+\Gamma\mu_{\mathrm{in}},  \label{Langsm}
\end{equation}
where $\mu=[a,b,c]^{T}$ are bosonic annihilation operators and their input noise operators are $\mu_{\mathrm{in}}=[a_{\mathrm{in}},b_{\mathrm{in}},c_{\mathrm{in}}]^{T}$ and $\Gamma=\rm{diag}[\sqrt{\kappa},\sqrt{\kappa},\sqrt{\kappa}]$ is the damping matrix. The coefficient matrix is
\begin{equation}
M=\left[\begin{array}{ccc}\omega -i\kappa   &  iJ\mathrm{e}^{-i\theta}+G   & iJ\mathrm{e}^{i\theta}+G  \\
 iJ\mathrm{e}^{i\theta}+G  & \omega -i\kappa    & iJ\mathrm{e}^{-i\theta}+G \\
 iJ\mathrm{e}^{-i\theta}+G  & iJ\mathrm{e}^{i\theta}+G  & \omega -i\kappa
\end{array}\right] . \label{Comatrix}
\end{equation}
the energy-levers are given by solving the eigenvalues of the coefficient matrix, i.e., $\left|M-\lambda I\right|=0$. Subsequently, we have
\begin{eqnarray}
\lambda^{'}_{1} &=&\omega -i\Gamma , \label{Aeigenhigh1} \\
\lambda^{'}_{\pm} &=&\omega -i\Gamma \pm \sqrt{\left( G^{2}-J^{2}\right) +iJG\left(\mathrm{e}^{i\theta}+\mathrm{e}^{-i\theta}\right)}.  \label{Aeigenhigh2}
\end{eqnarray}
Here, $\kappa$ represents the total loss of each mode. The eigenvalues in Eq.~(\ref{Aeigenhigh1}) and (\ref{Aeigenhigh2}) indicate the possibility of three-order exceptional points appearing.
\begin{figure}[t]
\includegraphics[scale=0.42]{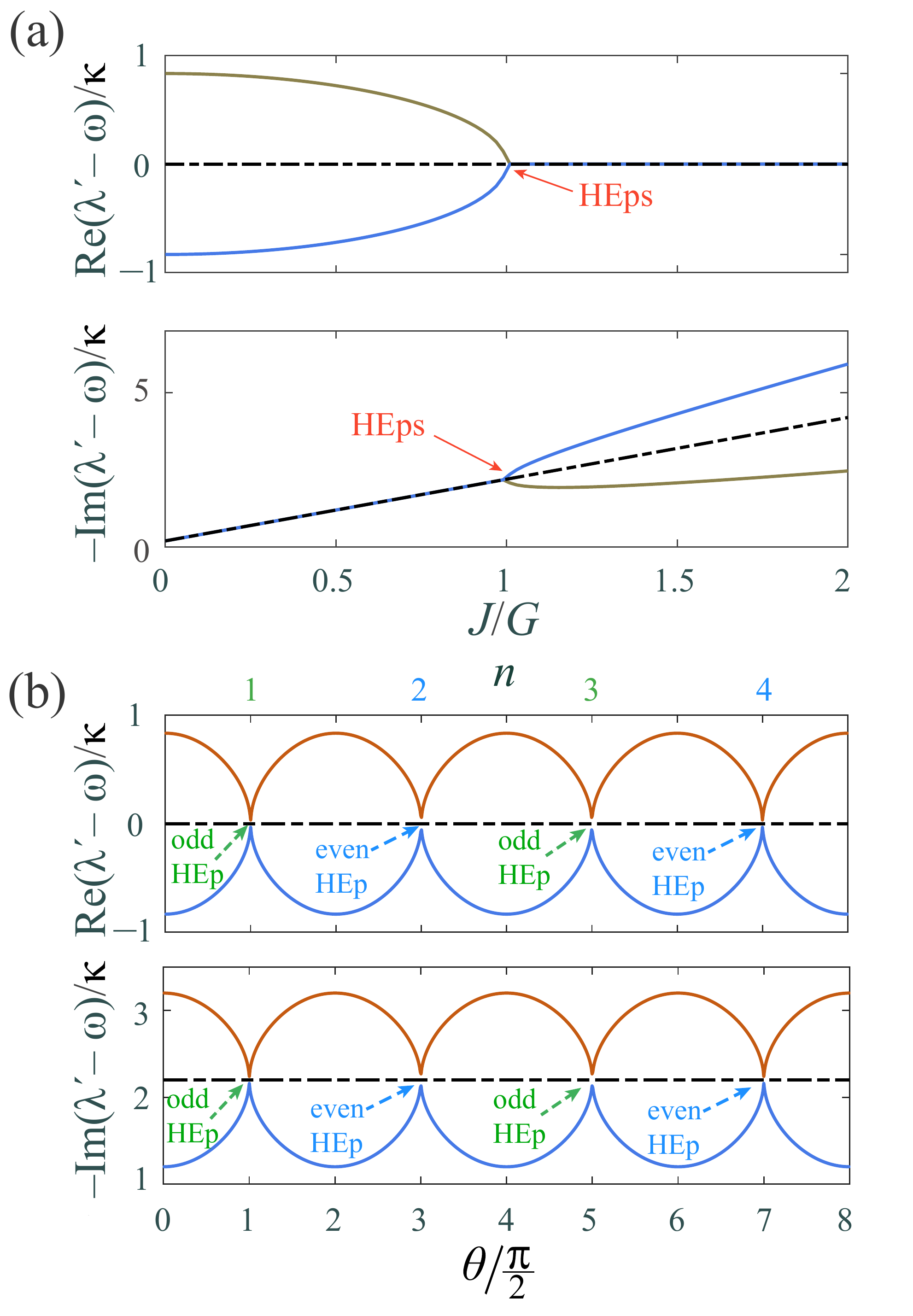}\centering
\caption{\textbf{(a)}, Eigenvalues as a function of dissipation coupling strength $J$  under phase-matching conditions, viz., $\theta/\frac{\pi}{2}=2n-1$ where $n\in\mathbb{Z}$. \textbf{(b)}, Eigenvalues as a function of parametric phase $\theta$ with balanced coherent and dissipative couplings, viz., $J=G$. The three-order exceptional points (abbreviated as HEPs) are marked by dashed arrows. The green (blue) arrows mark the odd (even) HEps with odd (even) phase-matching factor $n$, respectively.}
\label{fig8}
\end{figure}
As shown in Fig.~\ref{fig8} (a), the eigenvalues are plotted as functions of dissipative coupling strength $J$ under phase-matching conditions. All eigenvalues collapse to a single value with balanced coherent and dissipative couplings. These three-order degenerate points are called high-order Eps (HEps). Subsequently, as shown in Fig.~\ref{fig8} (b), the eigenvalues are plotted as a function of parametric phase $\theta$ with $J=G$. Clearly, the eigenvalues coincide, and HEps occur only under phase matching conditions, i.e., satisfying $\theta/\frac{\pi}{2}=2n-1, n\in \mathbb{Z}$.

Solving the Langevin equations~(\ref{Langsm}) in the frequency domain, we have
\begin{eqnarray}
\mu(\omega)=i(M-\omega)^{-1}\Gamma\mu_{in}(\omega).\label{Eq:muf}
\end{eqnarray}
Substituting Eq.~(\ref{Eq:muf}) into the input-output relation, we can obtain
\begin{eqnarray}
\mu_{out}(\omega)&=&\mu_{in}(\omega)-\Gamma\mu(\omega)\nonumber\\
&=&[I-i\Gamma(M-\omega I)^{-1}\Gamma\mu_{in}(\omega)\nonumber\\
&=&S(\omega)\mu_{in}(\omega),
\end{eqnarray}
where the transmission matrix $S(\omega)$ is given by
\begin{eqnarray}
S(\omega)=I-i\Gamma(M-\omega I)^{-1}\Gamma.
\end{eqnarray}
Defining $X=-[\kappa+i\delta]$, $J_1=J\mathrm{e}^{-i\theta}-iG$, and $J_2=J\mathrm{e}^{i\theta}-iG$, we can obtain
\begin{small}
\begin{equation}
S(\omega)=\frac{\kappa}{\Lambda}\left(
\begin{array}{cccc}
X^2-J_1J_2+1 & J_2^2-J_1X & J_1^2-J_2X\\
J_1^2-J_2X & X^2-J_1J_2+1 & J_2^2-J_1X\\
J_2^2-J_1X & J_1^2-J_2X & X^2-J_1J_2+1\\
\end{array}
\right), \label{FullM}
\end{equation}
\end{small}
where $\Lambda=X^3+J_1^3+J_2^3-3XJ_1J_2$. Hence, the transmission coefficient from mode $i$ to mode $j$ is given by the element $|S_{ji}|$. We have
\begin{eqnarray}
|S_{21}| &=&|S_{32}|=|S_{13}|=|\frac{\kappa}{\Lambda }(J_{1}^{2}-J_{2}X)| \\
&=&\big{|}\frac{\kappa \{(J\mathrm{e}^{-i\theta}-iG)^{2}+(J\mathrm{e}^{i\theta}-iG)[\kappa
+i(\Delta -\omega )]\}}{X^{3}+J_{1}^{3}+J_{2}^{3}-3XJ_{1}J_{2}}\big{|},  \nonumber
\end{eqnarray}%
\begin{eqnarray}
|S_{12}| &=&|S_{23}|=|S_{31}|=|\frac{\kappa}{\Lambda }(J_{2}^{2}-J_{1}X)| \\
&=&\big{|}\frac{\kappa \{(J\mathrm{e}^{i\theta}-iG)^{2}+(J\mathrm{e}^{-i\theta}-iG)[\kappa
+i(\Delta -\omega )]\}}{X^{3}+J_{1}^{3}+J_{2}^{3}-3XJ_{1}J_{2}}\big{|}.  \nonumber
\end{eqnarray}

Obviously, the parity of integer $n$ also affects energy transfer in between this three-mode system. At the odd HEps (corresponding to odd phase-matching factor $n$), the Hamiltonian becomes,
\begin{equation}
H =\left( \omega -i\kappa \right)\left(a^{\dag }a+b^{\dag}b+c^{\dag }c\right) + 2G\left(a^{\dag}b +b^{\dag}c+c^{\dag }a\right).   \label{HthreePodd}
\end{equation}
The cascade-coupling terms in the above Hamiltonian indicate circular energy transfer. At these odd HEps, energy transfers counterclockwise along $a\Rightarrow c \Rightarrow b \Rightarrow a$. In comparison, at these even HEps (corresponding to even phase-matching factor $n$), the system Hamiltonian becomes
\begin{equation}
H =\left(\omega -i\kappa \right)\left(a^{\dag}a+b^{\dag}b+c^{\dag}c\right) + 2G\left(b^{\dag}a +c^{\dag}b+a^{\dag}c\right). \label{HthreePeven}
\end{equation}
The energy transfers clockwise along $a\Rightarrow b \Rightarrow c \Rightarrow a$.

\begin{figure}[t]
\includegraphics[scale=0.42]{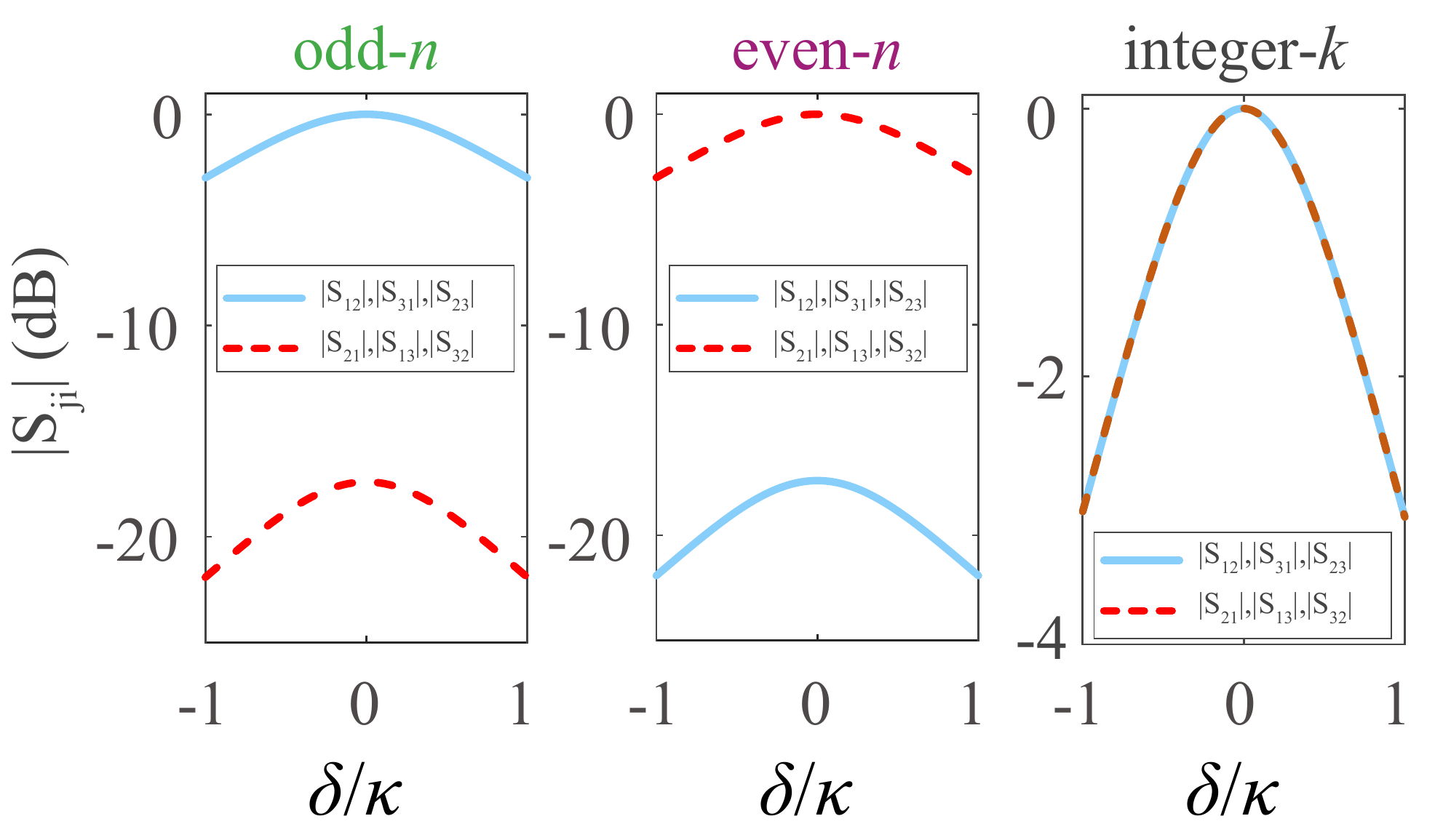}\centering
\caption{$|S_{\rm{ji}}|$ as a function of detuning $\delta$. At odd HEps with odd $n$, photo transfer counterclockwise. In comparison, at even HEps with even $n$, photons transfer clockwise. When the phases are not matched, for example, when $\theta/\frac{\pi}{2}=2k$ and $k\in\mathbb{Z}$, clockwise and counterclockwise energy transfer are exactly the same. The coherent coupling strengths are fixed at $G/\kappa_{\mathrm{i}}$=10 and the total decay rates for each cavity mode are set at $\kappa/J=100$ for the above calculations.}
\label{fig9}
\end{figure}

Figure~\ref{fig9} shows the transmission coefficient $|S_{ji}|$ (from port $i$ to port $j$) at odd or even HEps. The device featured a parity-dependent circulator: (i) energy transfers counterclockwise (i.e., port $2\Rightarrow 1$, port $1\Rightarrow3$, and port $3\Rightarrow2$) at odd HEps; (ii) energy transfers clockwise (i.e., port $2\Rightarrow 3$, port $3\Rightarrow1$, and port $1\Rightarrow2$) at even HEps. Around the HEps, energy transfer clockwise and counterclockwise is still nonreciprocal but with smaller isolation. Nevertheless, at $\theta/\frac{\pi}{2}=k$, where $k\in\mathbb{Z}$, clockwise and counterclockwise energy transfers are exactly the same, and the nonreciprocal property disappears.

Nonreciprocal microwave transport~\cite{de2020nonreciprocal} and directional amplification~\cite{de2019realization} have been realized in multimode superconducting cavity electromechanical devices. Optical nonreciprocal transmission has been experimentally demonstrated in silica microtoroid~\cite{ruesink2016nonreciprocity,ruesink2018optical, PhysRevLett.126.123603} and optomechanical crystal~\cite{fang2017generalized} multimode cavity optomechanical devices. These are ideal platforms for the experimental realization of our proposal. The coherent coupling between the cavity modes could be tuned by changing the distance between the fiber waveguide and the optical microcavity~\cite{ruesink2016nonreciprocity,ruesink2018optical, PhysRevLett.126.123603} or by changing the coupling capacitance (mutual inductance) of superconducting circuits~\cite{de2020nonreciprocal,de2019realization}. The key to the experimental realization is to achieve the inversion of the dissipation rate between the optical cavity and the mechanical modes. Recent experimental studies on non-Hermitian cavity optomechanics have shown that the decay rates of aluminum drums~\cite{toth2017dissipative}, silica microspheres~\cite{Shen2021dissipative}, and mechanical resonators can be much larger than the decay rate of each corresponding cavity resonator operating in the microwave or optical frequency bands. Through laser injection of the drum resonator~\cite{li2019controllable}, or exploitation of the gigahertz-frequency travelling acoustic wave mode~\cite{PhysRevLett.126.123603}, the decay rates of the mechanical resonators could be further increased to serval megahertz, which is dozens of times the dissipation rate of the corresponding microwave and optical cavity modes. Engineering the decay rates of these mechanical oscillators allows the experimental realization of our proposal in both the microwave and optical frequency domains.

\section{Conclusion}
In this work, we demonstrate broadband nonreciprocal and chiral photon transmission in a \textit{reversed-dissipation} cavity optomechanical system. We utilize a highly dissipative mechanical mode to endow the cavity modes with dissipative coupling with two kinds of phase factors. By steering the relative phase between the tunable parametric and fixed dissipative phases, destructive interference can occur between the coherent and dissipative couplings and the energy level exhibits \textit{periodic} Riemann-sheets with odd and even Eps.

At these Eps, the destructive interference breaks both $\mathcal{T}$- and $\mathcal{P}$-inversion symmetries. Subsequently, the photon transmission exhibits \textit{nonreciprocity} and \textit{chirality}. Moreover, reversing the dissipations of optical and mechanical modes allows the nonreciprocal and chiral bandwidth to be dictated by the cavity linewidth, which makes the broadband on-chip isolator accessible. Notably, the direction for unidirectional photon transfers in our proposal is determined by the parity of the phase-match factor $n$. Odd and even Eps result in exactly opposite directions, so they can be reversed by adjusting the parity of the phase factor $n$. Extending the scheme to a multiresonator coupled system, three-order Eps are observed, and a parity-dependent circulator can be constructed.

Our proposal also provides an \textit{in-situ} tunable parametric phase in addition to controlling the detunings or decay (gain) rates of the cavity modes~\cite{PhysRevA.95.013843, PhysRevA.96.023812,zhang2021exceptional, zhao2020observation,jiang2021energy,yu2019prediction,PhysRevA.104.012218,PhysRevB.103.235110,PhysRevB.102.161101,peng2016chiral, PhysRevA.104.063508, xu2016topological,doppler2016dynamically,zhang2019dynamically, zhang2017observation, peng2014loss} to control the photon transfer. The revealed cooccurrence of nonreciprocity and chirality will lead to the new direction of research and development of on-chip optical rectifier devices such as topology-protected photon diodes and circulators with new functionalities. Our findings have important implications for quantum information such as building directional quantum interfaces, cascaded quantum networks, and topological-protected state transfer and storage. Finally, the chirality associated with this scheme is also an important concept in modern physics. In quantum optics, with the rapid development of technologies~\cite{lodahl2015interfacing,gao2018programmable, roushan2017chiral}, many essential and intriguing quantum many-bodies arise from chiral interfaces and bring about the research field of chiral quantum optics~\cite{lodahl2017chiral}. Hence, this scheme has the potential to study quantum many-body physics that are unconventional from the perspective of condensed-matter physics.

\section{Supplementary Materials }
Supplementary material associated with this article can be found in the online version.

\section{Declaration of competing interest}
The authors declare that they have no conflicts of interest in this work.

\section{ACKNOWLEDGMENTS}
Z.~Chen was supported by the China Postdoctoral Science Foundation under Grant No. 2021M700442, and Y.~L.~Liu acknowledges the support of the Natural Science Foundation of China (NSFC) under Grant No. 12004044. H.~F.~Y acknowledges the support from the NSFC of China (Grants No.11890704), and the NSF of Beijing (Grant No.Z190012). T.~F.~Li acknowledges the support of the Development Program of China (Grant No. 2016YFA0301200), the National Natural Science Foundation of China (Grant No.62074091, and U1930402), the Science Challenge Project (Grant No.TZ2018003), and Tsinghua University Initiative Scientific Research Program.

Z.~Chen and Q.~Liu contributed equally to this work and share the first authorship.

 \bibliographystyle{elsarticle-num}





\end{document}